\documentclass[a4paper,12pt]{article}
\usepackage{graphicx}
\newtheorem{theo}{Theorem}[section]
\newtheorem{defi}{Definition}[section]
\newtheorem{exa}{Example} [section]

\newtheorem{rem}{Remark}[section]

\begin{document}

\par
\noindent
{\bf AN ALGORITHM FOR DETECTING HOPF-BIRFUCATION VARIETIES 
OF NONLINEAR POLYNOMIAL SYSTEMS 
\rm} 
\vskip 15 pt
\noindent
{Stelios Kotsios\\
      Faculty of Economics,\\
   Department of Mathematics and Computer Science, \\
University of Athens\\
         Pesmazoglou 8, Athens 10559, Greece\\
         {\tt \small skotsios@econ.uoa.gr}}
        
\date{}
\vskip 5 pt
\noindent
{\bf Abstract:}
An algorithm is presented here, for discovering Hopf-Bifurcation varieties of polynomial dynamical systems. It is based on the expression of specific polynomials, as sums of products of first degree polynomials, with parametrical coefficients. By giving to these parameters certain values, we ensure the positiveness of some quantities, constructing thereby proper Lyapunov functions, which guarantee the stability of the equilibrium point. The points where the afore mentioned  positiveness fails, define the Hopf-Bifurcation varieties upon discussion.
\section{Introduction}
It is widely known that when engineers and economists have to analyze mechanical, electrical or economical  dynamic phenomena, they are usually dealing with nonlinear dynamical systems. They give attention to the stability of those systems and to some special dynamic properties they possess (the existence of limit cycles, for instance), \cite{kn:Wiggins}. Another issue of practical importance is whether the systems maintain their dynamic behavior as certain parameters are varied. Especially, it is interesting if a given equilibrium point remains stable or unstable. These parameters are called bifurcation or Hopf-Bifurcation parameters and the values at which changes occur, if any, are called Hopf-Bifurcation points, \cite{kn:Guckenheimer},\cite{kn:Kuznetsov}.
\par
\noindent
The aim of this paper is to present an algorithm which detects Hopf-Bifurcation points of autonomous polynomial dynamical systems of the form:
\begin{equation}\label{In1}
\dot{\bf x}={\bf \Phi}({\bf x},{\bf \mu}) 
\end{equation}
where, ${\bf x}$ is the state vector, consisting from functions of $t$, ${\bf \Phi}$ is a vector function, consisting from polynomial functions of elements of ${\bf x}$ and ${\bf \mu}$ a set of parameters. Actually, the said algorithm, discovers the so called bifurcation varieties or bifurcation curves of a given equilibrium point of (\ref{In1}). These are sets of values of the parameters which satisfy certain relations. The violence of those relations implies the change of the stability behavior of the system around the equilibrium point.
\par
\noindent
There are a lot of efforts in the literature, for the description of proper algorithms which help us to determine bifurcation points. Let me refer to \cite{kn:Hu},\cite{kn:Dobson},\cite{kn:Wan}, to mention but a few.
The algorithm presented here works as follows. First it accepts a polynomial function $L$ as a Lyapunov function candidate, for the system (\ref{In1}), and an equilibrium point ${\bf x}_0$. Then it calculates the derivative of $L$ across the trajectories of (\ref{In1}), denoted by $\dot{L}$. After that, it checks the positiveness of the quantities $L$ and $-\dot{L}$. If this is true, under the assumption that the parameters satisfy a certain set of relations denoted by $B$, then we have stability of the equilibrium point and the boundary of $B$ defines the Hopf-Bifurcation variety. Indeed,  " crossing " this boundary it means that the relations $B$ do not hold any more and the stability of ${\bf x}_0$ collapses. 
\par
\noindent
To explore the positiveness of the quantities $L$ and $-\dot{L}$, we decompose them as follows:
\[ V=c_{1}(W_{i,\sigma,\varphi})[W_{1,-1,1}+ x_1]^{j_{1,1}} \cdot
[W_{2,-1,1}+W_{2,1,1}x_1+x_2]^{j_{2,1}} \cdot [W_{3,-1,1}+W_{3,1,1} x_1+W_{3,2,1} x_2 +x_3]^{j_{3,1}}
\cdots \]
\[\cdots [W_{n,-1,1}+W_{n,1,1}x_1 +W_{n,2,1}x_2+ \cdots +x_n]^{j_{n,1}}+ \]
\[ +c_{2}(W_{i,\sigma,\varphi})[W_{1,-1,2}+ x_1]^{j_{1,2}} \cdot
[ W_{2,-1,2}+W_{2,1,2}x_1+x_2]^{j_{2,2}} \cdot [W_{3,-1,2}+W_{3,1,2} x_1+W_{3,2,2} x_2 +x_3]^{j_{3,2}}
\cdots \]
\[\cdots [W_{n,-1,2}+W_{n,1,2}x_1 +W_{n,2,2}x_2+ \cdots +x_n]^{j_{n,2}}+ \cdots \]
\[+c_{k}(W_{i,\sigma,\varphi}) [W_{1,-1,k}+x_1]^{j_{1,k}} \cdot
[ W_{2,-1,k}+W_{2,1,k}x_1+x_2]^{j_{2,k}} \cdot [W_{3,-1,k}+W_{3,1,k} x_1+W_{3,2,k} x_2 +x_3]^{j_{3,k}}
\cdots \]
\begin{equation}\label{intro}
\cdots [W_{n,-1,k}+W_{n,1,k}x_1 +W_{n,2,k}x_2+ \cdots +x_n]^{j_{n,k}}+ R_{\cal W}
\end{equation}
where the exponents $j_{a,b}$ are specific positive whole numbers. The quantities
$W_{i,\sigma,\varphi}$ are undetermined parameters that can take real values,
 the coefficients $c_j(W_{i,\sigma,\phi})$ are depending on the parameters
$W_{i,\sigma,\phi}$ and the quantity $R_{\cal W}$ is a polynomial of the parameters $W_{i,\sigma,\phi}$ only, 
called the remainder. We obtain this " factorization " of the polynomials by means of a recursive algorithm, introduced in \cite{kn:hercma}, which resembles to the Euclidean Algorithm, and annihilates successively the maximum terms. 
Then, we seek for those values of the parameters $W_{i,\sigma,\phi}$, which eliminate the non-square terms and make the coefficients  of the square terms and the remainder, positive.
Obviously, if this can be achieved, the positiveness of $L$ and $-\dot{L}$ is secured.
\par
\noindent
The main advantages of the method are:
\par
1) Our approach is symbolic in nature and not numeric.
\par
2) The calculations can be easily carried out, since the coefficients in the expression (\ref{intro}) have a specific construction. Each of them contains a number of parameters which is larger or equal than the number of the parameters of the previous coefficient. This triangular structure, more known as a sparse system of algebraic equations, permits their easier handling, \cite{kn:strummfel}.
\par
3) The bifurcation values of the parameters ${\bf \mu}$, can be found straightforward, since they are involved in the calculations together with the artificial parameters $W_{i,\sigma,\phi}$, and therefore can be considered as polynomial functions of them.
\par
\noindent
We have to make clear that our method does not provide necessary and sufficient conditions. In other words, if our approach fails this does not mean that there are not bifurcation varieties or that another method could not find them.
\par 
\noindent
Throughout this paper ${\bf R}$ will denote the set of real numbers.

\section{Preliminaries}
In this section we present the basic tools on which the basic algorithm is relied.
 Let $(x_1,x_2,\ldots,x_n)$ be a vector of $n$-indeterminates which abbreviate by ${\bf x}$. An
expression of the form $ p=\sum_{\lambda=1}^{\varphi} c_{\lambda} x_1^{a_{1,\lambda}}
x_2^{a_{2,\lambda}} \cdots  x_n^{a_{n,\lambda}}$, where $c_{\lambda} \in {\bf R}$ and
some of the exponents $a_{i,j} \in \mathbf{Z}^+$ are not equal to zero, is called a
polynomial in $x_1,x_2,\ldots,x_n$ with real coefficients or, for short, a real
polynomial. An element $x_1^{a_{1,\lambda}} x_2^{a_{2,\lambda}}
\cdots  x_n^{a_{n,\lambda}}$ is called a {\it monomial} and an element $c_{\lambda}
x_1^{a_{1,\lambda}} x_2^{a_{2,\lambda}} \cdots  x_n^{a_{n,\lambda}}$ is called a {\it
term}. The quantity $c_{\lambda}$ is the coefficient of the term. The sum $a_{1,\lambda}+a_{2,\lambda}+\cdots+ a_{n,\lambda}$ is called {\it degree} of the term. A term is called {\it even} if all of its exponents are even integers, otherwise it is called an {\it odd} term. The term which corresponds to the exponent $(0,0,\ldots,0)$, is the constant term.
If we use the multi-index notation ${\bf a}_\lambda \in ({\bf Z}^+)^n$ to denote  the vector ${\bf a}_\lambda =(a_{1,\lambda},a_{2,\lambda},\ldots,a_{n,\lambda})$, we write a monomial compactly as ${\bf x}^{{\bf a}_\lambda}$ and a polynomial as $p=\sum_{\lambda=1}^{\varphi} c_{\lambda}{\bf x}^{{\bf a}_\lambda}$.
 The set of all real polynomials in $x_1,x_2,\ldots,x_n$ is denoted by
${\bf R}[x_1,x_2,\ldots,x_n]$ or ${\bf R}[{\bf x}]$. Let $\phi_{n,\lambda}= x_1^{a_{1,\lambda}} x_2^{a_{2,\lambda}} \cdots
x_n^{a_{n,\lambda}}$ and $\phi_{m,\mu}= x_1^{a_{1,\mu}} x_2^{a_{2,\mu}} \cdots
x_m^{a_{m,\mu}}$ be two monomials. We define the {\it lexicographical } order among
monomials \cite{kn:cox}, as follows: we say that $\phi_{n,\lambda}$ is ordered less
than $\phi_{m,\mu}$, denoted by $\phi_{n,\lambda} \prec \phi_{m,\mu}$, if either $n<
m$ or $n=m$ and in the vector difference $\phi_{n,\mu}-\phi_{m,\lambda}$ the left-most nonzero entry is positive. In other words, the monomials are ordered
as follows: $ x_1\prec \cdots \prec x_1^7 \prec$ $ \cdots \prec x_1x_2 \prec$ $ \cdots
\prec x_1 x_2^8 \prec$ $ \cdots \prec x_1x_2x_3 \prec \cdots$ Let $p$ be a given
polynomial, ordered lexicographically, the term that corresponds to the maximum
monomial is called the {\it maximum term} denoted by $maxterm(p)$, its degree is called the polynomial degree and it is denoted by $\deg(p,{\bf x})$.
\par
The next definitions will play a crucial role in the subsequents.
\begin{defi}
Let $\pi_i({\bf x)}$, $i=1, \ldots,m$ be a collection of polynomials in ${\bf R[x]}$. Then we set
\[V=V(\pi_i)=\{{\bf \theta} \in {\bf R}^n: \pi_i({\bf \theta})=0, \mbox{ for all } i=1,\ldots,m \}\]
We call $V$ the {\bf variety} defined by $\pi_i({\bf x)}$, $i=1, \ldots,m$.
\end{defi}

\begin{defi}
Let $\pi_i({\bf x)}$, $i=1, \ldots,m$ be a collection of polynomials in ${\bf R[x]}$. Then we set
\[A=A(\pi_i)=\{{\bf \theta} \in {\bf R}^n: \pi_i({\bf \theta})\le 0, \mbox{ for all } i=1,\ldots,m \}\]
We call $A$ the {\bf semi-algebraic set} defined by $\pi_i({\bf x)}$, $i=1, \ldots,m$.
\end{defi}
\begin{defi} Let $A$ be a {semi-algebraic set} defined by the polynomials $\pi_i({\bf x)}$, $i=1, \ldots,m$. The variety $\{{\bf \theta} \in {\bf R}^n: \pi_i({\bf \theta})=0\}$ is called the {\bf boundary} of $A$ and it is denoted by $\partial A$.
\end{defi}
Obviously, if we reverse the sense of the inequalities, nothing will change in the meaning of the above definitions.
\par
\noindent
Let us have a polynomial $p=\sum_{\lambda=1}^{\varphi}c_\lambda({\bf q}){\bf x}^{{\bf a}_\lambda}$, where the coefficients are polynomial expressions of a set of certain parameters ${\bf q}=(q_1,q_2,\ldots,q_k)$. Let us further suppose that we have the variety:
\[ U=\{ {\bf r}=(r_1,r_2,  \ldots,r_k) \in {\bf R}^k: \quad \pi_i({\bf r})=0, \quad i=1,\ldots,\theta \} \subset {\bf R}^k \]
where $\pi_i$ is a collection of polynomials of ${\bf q}$. We say that the polynomial $p$ is evaluated over the variety $U$, thus writing $p|_U$, if the parameters ${\bf q}$ take values from the set $U$. Rigorously, we have:
\[ p|_U=\left\{\sum_{\lambda=1}^{\varphi}c_\lambda({\bf r}){\bf x}^{{\bf a}_\lambda}, \quad {\bf r}\in U \right\} \]
If $U$ is a finite set then $p|_U$ is a finite set too, consisting from polynomials of ${\bf x}$, with real coefficients. If $U$ is an infinite variety, then $p|_U$ is an infinite set. 
If, furthermore, in this case a parametrization for the description of the variety is available, we can use it for the description of the set
$p|_U $, too, in a natural way.
Indeed, let us assume that we have the real functions
$\varphi_k:{\bf R}^{\lambda} \to {\bf R}$, $\lambda<\xi-1$ and that the points given by the relations
\[ r_k=\varphi_k(t_0,t_1,\ldots,t_{\lambda}), \quad \lambda<\xi-1, \quad k=0,\ldots,\xi-1, 
t_0,\ldots,t_\lambda \in {\bf R} \]
lie in $U$. These functions constitute a parameterization for the variety $U$ and the parameters ${\bf q}$. Then,
\[ p|_U=\left\{ \sum_{\lambda=0}^{\zeta}c_\lambda(\varphi_1(t_0,t_1, \ldots, t_\lambda),\ldots,\varphi_k(t_0,t_1, \ldots, t_\lambda)){\bf x}^{{\bf a}_\lambda} \right\} \]
The above terminology can be extended to a set ${\cal L}$ of polynomials, as follows:
\[{\cal L}|_U=\bigcup_{p\in {\cal L}}p|_U\]
Let us now suppose that we have $n$ functions of $t$ : $x_1(t)$, $x_2(t)$, $\ldots$, $x_n(t)$ and $n$ polynomial functions of $x_1,x_2,\ldots,x_n$, i.e., $\Phi_1(x_1,x_2,\ldots,x_n)$, $\Phi_2(x_1,x_2,\ldots,x_n)$, $\ldots$, $\Phi_n(x_1,x_2,\ldots,x_n)$. An expression of the form:
\[ \dot{x}_1=\Phi_1(x_1,x_2,\ldots,x_n)\]
\begin{equation}\label{ds0}
\dot{x}_2=\Phi_2(x_1,x_2,\ldots,x_n)
\end{equation}
\[ \vdots \]
\[ \dot{x}_n=\Phi_n(x_1,x_2,\ldots,x_n)\]
where by $\dot{x}_k$ we denote the derivative of the $x_k$ function, with respect to the time, it is called an autonomous polynomial dynamical system. We write it compactly as
$ \dot{\bf x}={\bf \Phi}({\bf x})$
where ${\bf x}(t)=(x_1(t),x_2(t),\ldots,x_n(t)) \in {\bf R}^n$, for each $t$, is the state space vector.
If the coefficients of the polynomials $\Phi_i$ depend on a certain set of parameters ${\bf \mu}=(\mu_1,\mu_2,\ldots,\mu_k)$ we write
\[ \dot{\bf x}={\bf \Phi}({\bf x,\mu})\]
A state ${\bf x}_0$ is called equilibrium point of (\ref{ds}), if and only if ${\bf \Phi}({\bf x}_0)=0$ or ${\bf \Phi}({\bf x}_0, {\bf \mu})=0$. In the latter case ${\bf x}_0$ may be depend on the parameters ${\bf \mu}$.
\par
\noindent
The next notions are classical in the literature, but we include them here for the sake of the self-reliance of the paper, \cite{kn:khalil}.
\begin{defi}
An equilibrium point ${\bf x}_0$ of (\ref{ds0}) is stable if for each $\epsilon>0$ there is $\delta=\delta(\epsilon)>0$ such that
\[ ||{\bf x}(0)||<\delta \Rightarrow ||{\bf x}(t)||<\epsilon, \quad \forall t \ge 0 \]
\end{defi}
\begin{defi}
The equilibrium point ${\bf x}_0 $ is unstable if it is not stable
\end{defi}
\begin{defi}
The equilibrium point ${\bf x}_0 $ is asymptotically stable if it is stable and $\delta$ can be chosen such that 
\[  ||{\bf x}(0)||<\delta \Rightarrow \lim_{t \to +\infty}||{\bf x}(t)||=0\]
\end{defi}
To determine stabilizability of a given equilibrium point, we adopt the Lyapunov function methodology. Let $L:D \to {\bf R}$ be a continuously differentiable function defined in a domain $D\subset {\bf R}^n$ that contains the equilibrium point. The derivative of $L$ along the trajectories of (\ref{ds0}), denoted by $\dot{L}({\bf x})$, is given by:
\[ \dot{L}({\bf x})=\sum_{i=1}^n\frac{\partial L}{\partial x_i} \dot{x}_i =
\sum_{i=1}^n\frac{\partial L}{\partial x_i}{\bf \Phi}(x_i) \]
We are ready now to establish the Lyapunov's stability theorem.
\begin{theo}\cite{kn:khalil}
Let ${\bf x}_0$ be an equilibrium point for (\ref{ds0}) and $D\subset {\bf R}^n $ be a domain containing ${\bf x}_0$. Let $L:D \to {\bf R}$ be a continuously differentiable function such that
\[ L(0)=0 \quad \mbox{and} \quad L({\bf x}) >0 \quad \mbox{in} \quad D-\{{\bf x}_0 \} \]
\[ \dot{L}({\bf x}) \le 0 \hspace{0.3cm} \mbox{in} \hspace{0.3cm} D \]
Then ${\bf x}_0$ is stable. Furthermore, if
\[ \dot{L}({\bf x}) <0 \hspace{0.3cm} \mbox{in} \hspace{0.3cm} D\]
 then ${\bf x}_0$ is asymptotically stable.
\end{theo}
\par
\noindent
The next definition introduces the notion of Hopf-Bifurcation variety (or Hopf-Bifurcation curve), which is a generalization of the well known Hopf bifurcation point. It is nothing else than a set of relations for the parameters ${\bf \mu}$ the violation of which can make the equilibrium point either stable or unstable, depending from the kind of violation. Specifically:
\begin{defi}
Let $\dot{\bf x}={\bf \Phi(x,\mu)}$ be an autonomous polynomial dynamical system depending from a set of parameters ${\bf \mu}=(\mu_1,\mu_2,\ldots,\mu_k)$ and ${\bf x}_0$ be an equilibrium point. Let, moreover, $\pi_j(\mu)$, $j=1,\ldots,\nu$, be a collection of polynomials of $\mu$. Let us have the semi-algebraic set
${\cal S}=\{ {\bf \theta} \in {\bf R}^k:\pi_j({\bf \theta})\le 0\}$. If ${\bf x}_0$ is asymptotical stable, for the system $\dot{\bf x}={\bf \Phi (x, \mu)}|_{\cal S}$ then the set $B=\partial {\cal S}$ is called the Hopf-Bifurcation variety of the parameters ${\bf \mu}$, at the point ${\bf x_0}$.
\end{defi}

\begin{exa}
Let us have the system
\[ \dot{x}=-yx+k x\]
\[ \dot{y}=\nu x^2\]
We write it compactly as $\dot{\bf x}={\bf \Phi}({\bf x}, {\bf \mu})$, with ${\bf x}=(x,y)$ and ${\bf \mu}=(k,\nu)$. This system has an infinite number of equilibrium points of the form $(0,\lambda)$, $\lambda \in {\bf R}$. If ${\cal S}$ and ${\cal M}$ are the varieties ${\cal S}=\{k=1,\nu=1\}$, ${\cal M}=\{(k,\nu): k=t,\nu=2t, t \in {\bf R} \}$, then the evaluation of the system over these varieties will give
\[ \dot{\bf x}={\bf \Phi}({\bf x , \mu})|_{\cal S} 
\Longleftrightarrow \begin{array}{l}
\dot{x}=-yx+x \\ \dot{y}=x^2 \end{array}\]
\[ \dot{\bf x}={\bf \Phi}({\bf x , \mu})|_{\cal M} 
\Longleftrightarrow \begin{array}{l}
\dot{x}=-yx+tx \\ \dot{y}=2tx^2 \end{array}\]
Let us now check if  $L=\nu x^2+(y-\lambda)^2$ is a Lyapunov function. Indeed, $L(0,\lambda)=0$ and $L({\bf x})>0$ for $\nu>0$. Furthermore, $\dot{L}({\bf x})=2\nu(k-\lambda)x^2$, which is negative or equal to zero if $k \le \lambda$. Therefore, the evaluation of the original system over the semi-algebraic set 
\[ {\cal S}=\{(k,\nu):k\le\lambda,\nu>0,\lambda \in {\bf R}\}\]
will provide us with a dynamical system, which is either stable or asymptotical stable (depending on the fact being $k=\lambda$ or $k < \lambda$) at the equilibrium point $(0,\lambda)$. The boundary of ${\cal S}$ is $\partial {\cal S}=\{ (k,\nu):k=\lambda, \nu=0, \lambda \in {\bf R}\}$ and it consists a bifurcation variety. 
\end{exa}
\section{The Algorithm}
\par

In this section we present the symbolic algorithm which discovers bifurcation varieties. This algorithm annihilates step by step the current maximum terms, by subtracting a suitable product of first degree polynomials with parametrical coefficients. We suppose that two algorithms are available, the first, named Solve-Algorithm, solves a system of polynomial equations $\{ p=0: p \in P\}$, the second, named InSolve-Algorithm, solves a system of inequalities $\{ p \le 0: p \in P \}$. The construction of such algorithms are the subject of  the research of computational algebra, and certain methodologies have been developed toward this scope, \cite{kn:cox},\cite{kn:cox2},\cite{kn:stetter}.
\begin{quote}
\vskip 10 pt \noindent \underline{\bf THE FORMAL-BIF-ALGORITHM} \vskip 10 pt \small {\sf
\par
\noindent {\bf Input:} The integer $m$,
the polynomials $\Phi_i({\bf x},{\bf \mu})$, $i=1, \ldots, l$,
the point ${\bf x_0}=(x_{0,1},x_{0,2},\ldots,x_{0,n})$, the sets of parameters ${\bf W}=\{ W_{i,\sigma,\varphi}\}$, 
${\bf S}=\{ S_{i,j,k}\}$ and ${\bf A}=\{ A_{i_1,i_2, \ldots, i_k} \}$
\vskip 6 pt
\noindent
{\bf Output:} The set $B$.
\vskip 6 pt
\noindent
{\bf Step 1:} Construct the polynomial ( Lyapunov Function Candidate).
\[ L=\sum_{k=1}^m\sum_{(i_1,i_2,\ldots,i_k) \in I_k\subset {{\bf Z}^+}^k}A_{(i_1,i_2,\ldots,i_k)}x_1^{i_1}x_2^{i_2}\cdots x_k^{i_k}\]
where $A_{(i_1,i_2,\ldots,i_k)}\in {\bf A}$. If $I_k=\emptyset$ , then the corresponding terms and their coefficients do not exist.
\par
\vskip 6 pt
\noindent
{\bf Step 2:} Use the Formal-Subroutine$[L,{\bf S}]$, that is with inputs the polynomial $L$ and the parameters ${\bf S}$, to obtain the sets $O_L,E_L$. The last element of $E_L$ is denoted by $R_L$.
\par
\vskip 6 pt
\noindent
{\bf Step 3:} Construct the polynomial
\[ V=-\sum_{i=1}^n\frac{\partial L}{\partial x_i}\Phi_i({\bf x},{\bf \mu})\]
\par
\vskip 6 pt
\noindent
{\bf Step 4:} Use the Formal-Subroutine$[V,{\bf W}]$, that is with inputs the polynomial $V$ and the parameters ${\bf W}$, to obtain the sets $O_V,E_V$.
\par
\vskip 6 pt
\noindent
{\bf Step 5:} Set $O=O_L\cup O_V\cup\{L(x_0)\}$, $E=E_L \cup E_V$, 
\par
\vskip 6 pt
\noindent
{\bf Step 6:} By means of the Solve-Algorithm, find the values of the parameters ${\bf A}$, ${\bf \mu}$, ${\bf S}$, ${\bf W}$ for which $\Theta=0$, for all $\Theta \in O$ . They construct a variety denoted by ${\cal O}$. 
\par
\vskip 6 pt
\noindent
{\bf Step 7:} Create the semi-algebraic set:
\[ J=\left\{ ({\bf A,\mu,S,W}) \in {\bf R}^{\varphi}: \Theta>0, \Theta \in(E_L-\{R_L\})|_{\cal O}, R_L|_{\cal O} \ge 0, H\ge 0, H \in E_V|_{\cal O} \right\}\]
 
\par
\vskip 6 pt
\noindent
{\bf Step 8:} By means of the InSolve-Algorithm check the feasibility of $J$, that is if $J\ne \emptyset$ or not.
\par
\vskip 6 pt
\noindent
{\bf Step 9:} {\bf IF} $J = \emptyset$ {\bf THEN} the method fails, {\bf stop} {\bf ELSE} denote by K the set of equations which defines the boundary $\partial J$ and then, set
\[ B=K-\{\Theta \in K, \quad \mbox{with}, \quad \deg (\Theta, {\bf \mu})=0 \} \]

Goto the output.
}
\par
\vskip 10 pt \noindent \underline{\bf THE FORMAL SUBROUTINE$[p,{\bf W}]$} \vskip 10 pt \small {\sf
\par
\noindent {\bf Input:} A set of undetermined parameters
${\bf W}=\{ W_{i,\sigma,\varphi}\}$, taking values in ${\bf R}$.
\par \noindent
 A multivariable polynomial 
 \[ p=\sum_{\lambda=1}^{\theta}c_\lambda({\bf \mu}) x_1^{a_{1,\lambda}}\cdots x_n^{a_{n,\lambda}}\quad,\quad a_{i,\lambda} \in {\bf N}\cup \{0\}, \quad i=1,\ldots,n, \quad \lambda=1, \ldots,\theta\]
 where the coefficients $c_\lambda({\bf \mu})$ depend from the vector of parameters
 ${\bf \mu}=(\mu_1, \ldots, \mu_\nu)$.
\par
\vskip 6 pt
\noindent
{\bf Output:} The sets $E_p$ and $O_p$. 
 
\par
\vskip 6 pt
\noindent
{\bf Initial Conditions:} $k=0$, $R_0=p$, $E_p=\{\}$, $O_p=\{\}$
\par
\vskip 6 pt
\noindent
{\bf REPEAT UNTIL} $R_k$ does not depend on any of the variables $x_1,x_2,\ldots,x_n$, {\bf THEN} set $E_p=E_p\cup\{R_k\}$.
\begin{description}
\item{\bf Step 1:} Set $k=k+1$.

\item{\bf Step 2:} Find the maximum term of $R_{k-1}$,
$maxterm(R_{k-1})=c_k({\bf \mu}, {\bf W}) x_1^{j_{1,k}}\cdots x_n^{j_{n,k}}$. The coefficient $c_k({\bf \mu}, {\bf W})$, in the
first iteration, is either a constant number or depends from the parameters ${\bf \mu}$ only. Then, it depends on the set of parameters
${\bf W}$, too.
\item{\bf Step 3:} {\bf IF} at least one of the exponents $j_{i,k}$, $i=1,\ldots,n$ is an odd positive integer {\bf THEN} $O_p=O_p\cup \{c_k({\bf \mu}, {\bf W})\}$
{\bf ELSE} $E_p=E_p\cup \{c_k({\bf \mu}, {\bf W})\}$
\item{\bf Step 4:} Form the polynomials:
\[\begin{array}{l}
 L_{1,k}= W_{1,-1,k}+x_1\\
 L_{2,k}=W_{2,-1,k}+W_{2,1,k}x_1+x_2\\
 \ldots \ldots\\
 L_{n,k}=W_{n,-1,k}+W_{n,1,k}x_1+W_{n,2,k}x_2+W_{n,3,k}x_3+\cdots +x_n
 \end{array}\]

\item{\bf Step 5:} Make the subtraction:
\[ R_k=R_{k-1}-c_k({\bf \mu}, {\bf W})L_{1,k}^{j_{1,k}} L_{2,k}^{j_{2,k}}\cdots L_{n,k}^{j_{n,k}}\]

\end{description}
 {\bf RETURN}}
 \end{quote}
\normalsize
The next two theorems describe the behavior of the Formal-Bif-Algorithm.
\begin{theo}
The Formal-Bif-Algorithm terminates, after a finite number of steps.
\end{theo}
{\bf Proof:} The termination of the algorithm depends from the termination of the Repeat procedure which appears at the Formal-Subroutine. But it can be easily proved that the operation at step 5 of the subroutine, annihilates the current maximum term of $R_{k-1}$ and thus, the maximum term of $R_k$ will be ordered less accordingly to the lexicographical order. Therefore, eventually, all the terms which contain at least one of the variables $x_1,x_2, \ldots,x_n$ will be eliminated and the procedure will stop.
\begin{theo}\label{amount}
The amount of the undetermine parameters $W_{ijk}$, appeared at the coefficient $c_k(\mu,{\bf W})$ of the step 5 of the subroutine, is larger or equal to the amount of parameters $W_{ijk}$, appeared at the coefficient 
$c_{k-1}({\mu},{\bf W})$.
\end{theo}
{\bf Proof:} Let $\rho x_1^{h_1}x_2^{h_2}\cdots x_n^{h_n}$ be the maximum term of the polynomial $p$. When we visit step 5 of the subroutine for first time, we get $C_1(\mu,{\bf W})=\rho$. This coefficient contains a zero amount of parameters $W_{ijk}$. The next higher ordered term is $\lambda x_1^{h_1-1}x_2^{h_2}\cdots x_n^{h_n}$. This term may exist at the polynomial $p$ but it will be also created by the product $L_{1,1}^{h_1}L_{2,1}^{h_2} \cdots L_
{n,1}^{h_n}$$=(W_{1,-1,1}+x_1)^{h_1}$$(W_{2,-1,k}+W_{2,1,k}x_1+x_2)^{h_2}$$\cdots$ of the step 4 of the subroutine. Using the binomial theorem (Newton's theorem) for the expression $(W_{1,-1,1}+x_1)^{h_1}$, we finally get:
$c_2(\mu,{\bf W})=\lambda -h_1 W_{1,-1,1}$. This coefficient contains obviously a larger amount of W-parameters than the previous one. For the coefficient $c_3(\mu,{\bf W})$ we have
$c_3(\mu,{\bf W})=g-h_2W_{1,-1,2}-\frac{h_2(h_2-1)}{2}W^2_{1,-1,1}$, which has the same property and therefore, working inductively, we can establish the theorem.
\vskip 8 pt
The next theorem is devoted to the efficiency of the algorithm to the discovering of the bifurcation varieties.
\begin{theo}
Let us suppose that we have the nonlinear dynamical system
\begin{equation}\label{ds}
\dot{\bf x}={\bf \Phi}({\bf x},{\bf \mu})
\end{equation}
where ${\bf \Phi}=(\Phi_1,\Phi_2,\ldots,\Phi_p)$ are polynomials of ${\bf x}$, with polynomials expressions of ${\bf \mu}$, as coefficients, and ${\bf x}_0$ an equilibrium point. We apply the Formal-Bif-Algorithm and let $B$ be its output. If $B \ne \emptyset$, then all the members of $B$ define the Hopf-Bifurcation variety of (\ref{ds}), at ${\bf x_0}$.
\end{theo}
{\bf Proof:} We shall follow the Formal-Bif-Algorithm step by step. Step 1 will create a Lyapunov function candidate. Step 3 will create the quantity $V$, which is the opposite of its derivative across the orbits of the system. Both of them are polynomials of ${\bf x}$ with parametrical coefficients. Applying the Formal-Subroutine we get the next expressions for ${ L}$ and ${ V}$ correspondingly:
\[{ L}=\sum_{r=1}^k\tilde{c}_{ r}\tilde{L}_{1,r}^{j_{1,r}}\tilde{L}_{2,r}^{j_{2,r}}\cdots \tilde{L}_{n,r}^{j_{n,r}}+R_L\]

\[{ V}=\sum_{r=1}^{k'}{c}_{ r}{L}_{1,r}^{j_{1,r}}{L}_{2,r}^{j_{2,r}}\cdots {L}_{n,r}^{j_{n,r}}+R_V\]
with
\[\begin{array}{l}
 \tilde{L}_{1,r}=S_{1,-1,r}+x_1\\
 \tilde{L}_{2,r}=S_{2,-1,r}+S_{2,1,r}x_1+x_2\\
\qquad  \vdots\\
\tilde{L}_{n,r}=S_{n,-1,r}+S_{n,1,r}x_1+S_{n,2,r}x_2+S_{n,3,r}x_3+\cdots +x_n
\end{array}\]
and
\[ \begin{array}{l}
L_{1,r}=W_{1,-1,r}+x_1\\
{L}_{2,r}=W_{2,-1,r}+W_{2,1,r}x_1+x_2\\
\qquad  \vdots\\
L_{n,r}=W_{n,-1,r}+W_{n,1,r}x_1+W_{n,2,r}x_2+W_{n,3,r}x_3+\cdots +x_n \end{array} \]
The coefficients $\tilde{c}_{ r},{c}_{ r}$ are polynomial expressions of the parameters ${\bf A}, {\bf W}, {\bf S}$ and ${\bf \mu}$.
We obtain these expressions by backward substitution of the values of $R_k$, obtained in the step 5 of the subroutine. Step 6, in combination with the construction of the set $O$, finds those values of the parameters ${\bf A,\mu,S}$ and ${\bf W}$, which guarantee that all the odd terms of ${ L}$ and $V$ will be eliminated. Furthermore they ensure that the Lyapunov function candidate is equal to zero at the equilibrium point ${\bf x}_0$. Step 7 of the algorithm, constructs a set of variables which make the coefficients of the even terms of the $L$ strictly positive, the remainder $R_L$ positive or equal to zero and the coefficients of the even terms of $V$ strictly positive or zero.
 All the above indicate that the above expression ${ L}$ is a Lyapunov function for the nonlinear system, which is negative across its orbits and thus ${\bf x}_0$ is stable or asymptotical stable depending if $V\le 0$ or $V<0$. 
Step 9 provides us with the boundary of the previous set, in other words this particular set which "separates" the two different " behaviors " those of stability and instability of the equilibrium point. The subset of the above set consisting only with equations which involves the parameters ${\bf \mu}$, (step 9), defines the bifurcation variety. 
 
 \par
\begin{rem}
The keynote of the whole process is the solvability of the polynomial equations or inequalities, appeared at the steps 6 and 7 of the algorithm. This is the classical problem in algebraic geometry and certain methods have been developed toward this direction. It still remains a hard problem. Nevertheless, the polynomials appeared at the sets $O$ or $J$ have a particular structure, which make the solution of the equations easier. Indeed, accordingly to theorem \ref{amount}, they have a "triangular" construction and can be faced via methods of solution devoted to sparse systems, \cite{kn:strummfel}.
\end{rem}

\section{Examples}
To clarify the previous algorithm we present certain examples.
\par \noindent
{\bf Example 1.}
To indicate how the algorithm works in practice, we deal firstly with a rather simple example. We consider the system
\[ \dot{x}=\mu x - x^3\]
\[ \dot{y}=-y\]
In this case  ${\bf x}=(x,y)$, ${\bf \mu}=\mu$ and ${\bf \Phi}=(\Phi_1,\Phi_2)$ with 
$\Phi_1=\mu x - x^3$ and $\Phi_2=-y$.
This system has two equilibrium points $(0,0)$ and $(\mu,0)$. By means of classical tools, that is the Jacobian, we can show that $(0,0)$ is a stable point for $\mu <0$ and a saddle point for $\mu>0$, and that $(\mu,0)$ is a stable point for $\mu>0$ and a saddle point for $\mu<0$, \cite{kn:khalil}. 
\par
\noindent
Let us handle this system by using the algorithm developed previously. We shall work with the point $(0,0)$. We take as $L$ a second degree homogeneous polynomial, that is 
$ L=A_1x^2+A_2y^2+A_3xy$, and we follow the Formal-Bif-Algorithm step by step. By substituting backwards the results of the step 5 of the subroutine we shall take the next expression for the polynomial $L$:
\[ L=A_2(S_{2,-1,1}+S_{2,1,1}x+y)^2+(A_3-2A_2S_{2,1,1})(S_{1,-1,2}+x)(S_{2,-1,2}+S_{2,1,2}x+y)+\]
\[+(-A_3S_{1,-1,2}-2A_2S_{2,-1,1}+2A_2S_{1,-1,2}S_{2,1,1})(S_{2,-1,3}+S_{2,1,3}x+y)+\]
\[+(A_1-A_2S_{2,1,1}^2-A_3S_{2,1,2}+2A_2S_{2,1,1}S_{2,1,2})(S_{1,-1,4}+x)^2+\cdots+R_L\]
with 
\[ R_L=-A_1S_{1,-1,4}^2+2A_1S_{1,-1,4}S_{1,-1,5}-A_2S_{2,-1,1}^2+\cdots\]
For the quantity $V=2A_1x^4-2\mu A_1x^2+2A_2y^2$$+A_3xy+A_3x^3y-\mu A_3xy$, we get:
\[ V=2A_2(W_{2,-1,1}+W_{2,1,1}x+y)^2+A_3(W_{1,-1,2}+x)^3(W_{2,-1,2}+W_{2,1,2}x+y)+\]
\[+(-3W_{1,-1,2}A_3)(W_{1,-1,3}+x)^2(W_{2,-1,3}+W_{2,1,3}x+y)+\]
\[+(A_3-\mu A_3-3A_3W_{1,-1,2}^2+6A_3W_{1,-1,2}W_{1,-1,3}-4A_2W_{2,1,1})(W_{1,-1,4}+x)(W_{2,-1,4}+W_{2,1,4}x+y)+\]
\[+c_1(W_{2,-1,5}+W_{2,1,5}x+y)+(2A_1-W_{2,1,2}A_3)(W_{1,-1,6}+x)^4+\]
\[+c_2(W_{1,-1,7}+x)^3+c_3(W_{1,-1,8}+x)^2+c_4(W_{1,-1,9}+x)+R_V\]
(We do not write the coefficients $c_1,c_2,c_3,c_4$ and the remainder $R_V$, explicitly due to their large size). Thus, the set $O$, 
consisting from all the coefficients of the odd terms of $L$ and $V$, is
\[ O=\{ A_3-2A_2S_{2,1,1},\hspace{0.2cm} -A_3S_{1,-1,2}-2A_2S_{2,-1,1}+2A_2S_{1,-1,2}S_{2,1,1}, \ldots, \hspace{0.2cm}A_3,\hspace{0.2cm}-3W_{1,-1,2}A_3,\]
\[
A_3-\mu A_3-3A_3W_{1,-1,2}^2+6A_3W_{1,-1,2}W_{1,-1,3}-4A_2W_{2,1,1},\ldots,0\}\]
The last number $0$, corresponds to the evaluation of the Lyapunov function candidate at the equilibrium point ${\bf x_0}=(0,0)$.
The sets $E_L$ and $E_V$, consisting from the coefficients of the even terms of $L$ and $V$, correspondingly, are:
\[ E_L=\{A_2, \hspace{0.2cm}A_1-A_2S_{2,1,1}^2-A_3S_{2,1,2}+2A_2S_{2,1,1}S_{2,1,2},\ldots, R_L\}\]
\[ E_V=\{ 2A_2,\hspace{0.2cm}2A_1-W_{2,1,2}A_3,\ldots,R_V\}\]
Now, the variety ${\cal O}$, consisting from the values of the parameters which vanish the non-square terms, is:
\[ {\cal O}=\{S_{2,-1,1}=0,\hspace{0.2cm}W_{2,-1,1}=0,\hspace{0.2cm} S_{1,-1,4}=0,\hspace{0.2cm}W_{2,1,1}=0,\hspace{0.2cm}S_{2,1,1}=0,\hspace{0.2cm}\]
\[W_{1,-1,6}=0,\hspace{0.2cm}W_{1,-1,8}=0,\hspace{0.2cm}A_3=0\}\]
Evaluating $E_L$ and $E_V$ over the values of ${\cal O}$, we get:
\[ E_L|_{\cal O}=\{A_1,A_2\}\qquad,\qquad E_V|_{\cal O}=\{2A_1,-2A_1\mu,2A_2\}\]
These values of the coefficients produce the next expressions for $L$ and $V$:
\[ L|_{\cal O}=A_1x^2+A_2y^2,\quad V|_{\cal O}=2A_1x^4-2A_1\mu x^2+2A_2 y^2\]
Now, the semi-algebraic set ${J}$ is
\[ { J}=\{ (A_1,A_2,{\bf \mu}), A_1>0,A_2>0,2A_1>0,-2A_1\mu>0,2A_2>0 \} \]
which is feasible for $A_1>0,A_2>0,\mu<0$. In other words, the above values guarantee the positiveness of $L$ and $V$, which ensure the stability of the origin. The boundary $\partial J$ is defined by the equations $A_1=0,A_2=0,\mu=0$. The only equation which involves the parameter is $\mu=0$, this equation defines the bifurcation variety $B=\{\mu:\mu=0\}$.
This result coincides with that provided by the classical theory, \cite{kn:khalil}. We can repeat a similar analysis for the point $(\mu,0)$, too.
\par
\noindent
{\bf Example 2.} In this example we exhibit the applicability of the method in the case of one dimension dynamical systems. Let us consider the system:
\[ \dot{x}=ax^3+bx^2+cx+d \]
To simplify the manipulation  we take $d=-(a+b+c)$. This means that $x=1$ is an equilibrium point. We shall work with this specific point. As Lyapunov function candidate we shall use the quantity $L=x^2+A_1x+1$. The Formal Bif-Algorithm will give:
\[ L=(x+S_{1,-1,1})^2+(A_1-2S_{1,-1,1})(x+S_{1,-1,2})+\]
\[+(1-S_{1,-1,1}^2-A_1S_{1,-1,2}+2S_{1,-1,1}S_{1,-1,2})\]
and
\[ V=-2a(x+W_{1,-1,1})^4+(-2b-aA_1+8aW_{1,-1,1})(x+W_{1,-1,2})^3+\]
\[+c_1(x+W_{1,-1,3})^2+c_2(x+W_{1,-1,4})+R_V \]
We seek those values of the parameters, which will eliminate the non-even terms. This is achievable if $b=-3a$, and thus ${\cal O}=\{$$S_{1,-1,1}=-1$, $A_1=-2$, $W_{1,-1,3}=-1$, $W_{1,-1,1}=-1\}$.
Evaluating $L$ and $V$ over ${\cal O}$ we take:
\[ L|_{\cal O}=(x-1)^2 \quad,\quad V|_{\cal O}=(6a-2c)(x-1)^2-2a(x-1)^4 \]
The positiveness of $V$ is guaranteed by the 
feasibility of the set $J=\{(a,c): 6a-2c<0, \quad -2a>0\}$. This set is indeed non-void and $\partial J$ is defined by the equations $6\theta-2c=0$, for any $\theta<0$ and $a=0$. Thus, the bifurcation variety is
$\{(a,c):a=0,c=3\theta\}$,for any given $\theta<0$.
\par
\noindent
{\bf Example 3.} To illustrate the compatibility of the method with linear systems, we examine the next case:
\[ \dot{x}=\alpha x+ \beta y \]
\[ \dot{y}=\gamma x + \delta y \]
The origin $(0,0)$ is the only equilibrium point. Let us work with the function $L=x^2+y^2$. This is a classical Lyapunov function and thus we shall not deal with it any more. For the quantity $V$ the Formal Bif-Algorithm will give:
\[ V=-2\delta (W_{2,-1,1}+W_{2,1,1}x+y)^2+(-2\beta-2\gamma+4\delta W_{2,1,1})(W_{1,-1,2}+x)(W_{2,-1,2}+\]
\[+W_{2,1,2}x+y)+W_{1,-1,2}(2\beta+2\gamma+4\delta-4\delta W_{2,1,1})\]
\[(W_{2,-1,3}+W_{2,1,3}x+y)+[-2\alpha+2\delta W_{2,1,1}(W_{2,1,1}-2W_{2,1,2})+2W_{2,1,2}(\beta+\gamma)]\]
\[\cdot (W_{1,-1,4}+x)^2+c_1(W_{1,-1,5}+x)+R_V\]
The next set of values of the parameters ${\cal O}=\{$$S_{2,-1,1}=0$, $S_{1,-1,4}=0$, $W_{2,-1,1}=0$, $W_{1,-1,4}=0$, $S_{2,1,1}=0$, $W_{2,1,1}=\frac{\beta+\gamma}{2\delta}\}$, will give:
\[ L|_{\cal O}=x^2+y^2\quad,\quad V|_{\cal O}=\left[-2\alpha+\frac{(\beta+\gamma)^2}{2\delta}\right]x^2-2\delta \left[y+\frac{x(\beta+\gamma)}{2\delta}\right]^2 \]
The set $J$ is $J=\{ (\alpha,\beta,\gamma,\delta): $$\delta<0$ and $\left[-2\alpha+\frac{(\beta+\gamma)^2}{2\delta}\right]>0\}$. This set is feasible and the equations
$\delta=0$, $\left[-2\alpha+\frac{(\beta+\gamma)^2}{2\theta}\right]=0$, with $\theta <0$ define the bifurcation variety upon request.

\par
\noindent
{\bf Example 4.} Let us have the system:
\[ \dot{x}=-(y-\alpha)x\]
\[ \dot{y}=\gamma x^2\]
where $\alpha$ and $\gamma$ are parameters taking values in ${\bf R}$. The equilibrium points are $(0,k)$, $k \in {\bf R}$. We shall try to find bifurcation varieties of the parameters for a specific equilibrium point $(0,\theta)$, $\theta$ is a constant but otherwise arbitrary real number. We shall work with the following  Lyapunov function candidate:
\[ L=A_1 x^2+A_2y^2+A_3x+A_4y+A_5 \]
The Formal-Bif-Algorithm will give for $L$:
\[L=A_2(S_{2,-1,1}+S_{2,1,1}x+y)^2+(-2A_2S_{2,1,1})(S_{1,-1,2}+x)(S_{2,-1,2}+S_{2,1,2}x+y)+\]
\[+(A_4-2A_2S_{2,-1,1}+2A_2S_{1,-1,2}S_{2,1,1})(S_{2,-1,3}+S_{2,1,3}x+y)+\]
\[+(A_1-A_2S_{2,1,1}^2+2A_2S_{2,1,1}S_{2,1,2})(S_{1,-1,4}+x)^2+c_1(S_{1,-1,5}+x)+R_L\]
For the quantity $V$, we get:
\[ V=(2A_1-2\gamma A_2)(W_{1,-1,1}+x)^2(W_{2,-1,1}+W_{2,1,1}x+y)+(A_3-4A_1W_{1,-1,1}+\]
\[+4\gamma A_2 W_{1,-1,1})(W_{1,-1,2}+x)(W_{2,-1,2}+W_{2,1,2}x+y)+(-2A_1W_{1,-1,1}^2+2\gamma A_2W_{1,-1,1}^2-\]
\[-A_3W_{1,-1,2}+4A_1W_{1,-1,1}W_{1,-1,2}-4\gamma A_2W_{1,-1,1}W_{1,-1,2})(W_{2,-1,3}+W_{2,13}x+y)+\]
\[+(-2A_1W_{2,1,1}+2\gamma A_2 W_{2,1,1})(W_{1,-1,4}+x)^3+h_1(W_{1,-1,5}+x)^2+h_2(W_{1,-1,6}+x)+R_V\]
and thus the sets $O$ and $E$ are:
\[O=\{-2A_2S_{2,1,1}, \hspace{0.2cm}A_4-2A_2S_{2,-1,1}+2A_2S_{1,-1,2}S_{2,1,1}, \hspace{0.2cm} c_1, \hspace{0.2cm} 2A_1-2\gamma A_2, \]
\[\hspace{0.2cm}A_3-4A_1W_{1,-1,1}+4\gamma A_2 W_{1,-1,1},\ldots, h_2, \hspace{0.2cm}\theta^2A_2+\theta A_4+A_5\}\]
\[E=\{A_2, \hspace{0.2cm} A_1-A_2S_{2,1,1}^2+2A_2S_{2,1,1}S_{2,1,2}, \hspace{0.2cm}h_1, \hspace{0.2cm}R_L,\hspace{0.2cm}R_V\}\]
(The coefficients $h_1,h_2$ are large polynomial expressions of the parameters $W_{ijk}$.) The nontrivial values of these parameters which make the members of the set $O$ equal to zero are:
\[{\cal O}=\{A_1=\gamma A_2, \hspace{0.2cm} A_4=2A_2S_{2,-1,1}, \hspace{0.2cm}A_5=-\theta^2 A_2-\theta A_4, \hspace{0.2cm}, S_{1,-1,4}=0,\]
\[A_3=0, \hspace{0.3cm} S_{2,1,1}=0, \hspace{0.3cm} W_{1,-1,5}=0\}\]
and the quantities $L$ and $V$ become
\[ L=A_2\gamma x^2 +A_2(S_{2,-1,1}+y)^2-A_2(\theta+S_{2,-1,1})^2\]
\[ V=-2\gamma A_2(\alpha +S_{2,-1,1})x^2\]
In order to ensure that the coefficients of the above polynomials are positive or equal to zero
the next set must be feasible.
\[ J=\{ (A_1,A_2,\gamma,\theta,S_{2,-1,1}): A_2 \gamma > 0, A_2 > 0, -A_2(\theta + S_{2,-1,1})^2 \ge 0, -2\gamma A_2(\alpha +S_{2,-1,1})\ge 0\]
The only non-trivial way to get that is:
\[ S_{2,-1,1}=-\theta, \hspace{0.2cm}A_2>0, \hspace{0.2cm}\gamma>0, \hspace{0.2cm} a+\theta<0\]
 and thus the bifurcation variety for the parameters $\alpha$ and $\gamma$ are $\{(\alpha,\gamma):\alpha=-\theta, \gamma:\gamma=0\}$. 
\par
\noindent
{\bf Example 5.} The current example deals with a nonlinear system with three states:
\[ \dot{x}=\varphi y^2-xy+ax+7y\]
\[ \dot{y}=x^2-\varphi xy+5x+\varphi y\]
\[ \dot{z}=-2z\]
The equilibrium points are $(0,0,0)$ and
\[ \left( \frac{-203-22a \varphi+a^2 \varphi^2\pm 6\sqrt{1764+455a \varphi+22a^2 \varphi^2-a^3 \varphi^3}}{(7+a \varphi)(13+a    \varphi)},\right.\]
\[\left.
\frac{-42 \mp \sqrt{1764+455a \varphi+22a^2 \varphi^2-a^3 \varphi^3}}{13 \varphi+a \varphi^2},0\right)\]
We shall work with the first one. We choose as Lyapunov function the classical one $L=x^2+y^2+z^2$. The Formal Bif-Algorithm will give for $V$:
\[V=4(W_{3,-1,1}+W_{3,1,1}x+W_{3,2,1}y+z)^2-8W_{3,2,1}(W_{2,-1,2}+W_{2,1,2}x+y)\cdot \]
\[\cdot (W_{3,-1,2}+W_{3,1,2}x+W_{3,2,2}y+z)-8(W_{3,1,1}-W_{2,1,2}W_{3,2,1})\cdot (W_{1,-1,3}+x) \cdot\]
\[\cdot (W_{3,-1,3}+W_{3,1,3}x+W_{3,2,3}y+z)-8(w_{3,-1,1}-W_{1,-1,3}W_{3,1,1}-W_{2,-1,2}W_{3,2,1}+\]
\[+W_{1,-1,33}W_{2,1,2}W_{3,2,1})(W_{3,-1,4}+W_{3,1,4}x+W_{3,2,4}y+z)-2(\varphi+2W_{3,2,1}^2-4W_{3,2,1}W_{3,2,2})\cdot\]
\[\cdot (W_{2,-1,5}+W_{2,1,5}x+y)^2+\sigma_1 (W_{1,-1,6}+x)(W_{2,-1,6}+W_{2,1,6}x+y)+\sigma_2(W_{2,-1,7}+W_{2,1,7}x+y)+\]
\[+\sigma_3(W_{1,-1,8}+x)^2+\sigma_4(W_{1,-1,9}+x)+R_V\]
(we do not write $\sigma_1,\sigma_2$,$\sigma_3$,$\sigma_4$,$R_V$ explicitly, due to their large size).
The values ${\cal O}=\{$ $W_{3,2,1}=0, W_{3,-1,1}=0$, $W_{2,-1,5}=0$, $W_{1,-1,8}=0$, $W_{3,1,1}=0$, $W_{2,1,5}=\frac{1}{\varphi}\}$ will give the next expressions for $L$ and $V$:
\[ L=x^2+y^2+z^2\]
\[V|_{\cal O}=\left(-2a+\frac{72}{\varphi}\right)x^2-2\varphi \left(y+\frac{6x}{\varphi}\right)^2+4z^2\]
Obviously,  for $\varphi<0$, $-2a+\frac{72}{\varphi}>0$ $V$ is positive and hence the Lyapunov function is decreasing across the trajectories of the system and the origin is stable. Therefore, the relations  $\varphi=0$ and $-2a+\frac{72}{\rho}=0, \rho<0$ define the bifurcation variety upon request.
\section{Concluding Remarks}
The issue of this paper was the description of an algorithm which finds the so-called Hopf-Bifurcation varieties. These are relations among parameters, the violation of which changes the stability of an equilibrium point. This algorithm transforms polynomial expressions to sums of products of first degree polynomials with parametrical coefficients. By giving to these parameters proper values we ensure the positiveness of certain quantities and thus, we can investigate the stability behavior through Lyapunov theory.

\end{document}